\documentclass[12pt]{article}
\sf

\title{Fast Recompilation of Object Oriented Modules}

\author{J\'er\^ome Collin and Michel Dagenais \\
\ \\
D\'epartement de g\'enie informatique \\
Ecole Polytechnique de Montr\'eal \\
C.P. 6079, Succ. Centre-Ville \\ 
Montr\'eal, Qu\'ebec, CANADA, H3C 3A7 \\
michel.dagenais@polymtl.ca \\
\  \\
}

\begin{document}

\maketitle
%

\begin{abstract}
Once a program file is modified,
the recompilation time should be minimized, without
sacrificing execution speed or high level object oriented features.
The recompilation time is often a problem
for the large graphical interactive distributed
applications tackled by modern OO languages. 
A compilation server and
fast code generator were developed and integrated with the SRC Modula-3
compiler and Linux ELF dynamic linker.
The resulting compilation and recompilation speedups are impressive.
The impact of different language features, processor speed,
and application size are discussed.
\end{abstract}


\section{Introduction}

Recompilation speed is only one ingredient in the global picture
of programming productivity. Nonetheless, its impact on programmer
satisfaction is not to be underestimated. Interpreters are
sometimes seen as the solution for zero recompilation time.
However, the modified module needs to be parsed into bytecode,
which is not too different from machine code, especially if the
link editing phase is partially obviated by dynamic linking.

The recompilation may be divided into the following phases.
The smart recompilation phase determines the modified files and
computes the minimal set of files to recompile.
The parsing and code generation phases convert the source code files
into relocatable binary files. The prelink phase computes package wide
information such as initialization order. Finally, the link phase combines
all the relocatable binary files into a library or executable program.

These phases are detailed in section 2 along with a discussion of the
impact of different language features on their complexity.
This section motivates the two main extensions brought by the authors 
to the DEC SRC Modula-3 compiler \cite{bk-m3sp}, 
a compilation server and fast code generator,
and reviews related work.

Section 3 details the compilation server while section 4
describes the fast integrated code generator for Linux ELF \cite{bk-abi386}. 
Section 5
presents the results obtained with the enhanced compiler, and outlines the
contribution of each extension as well as the sensitivity to different
parameters. In the conclusion, the applicability of these results to other
languages such as Java \cite{JavaLanguage} and C++ \cite{bk-cpp}
are discussed, and avenues for further development
are examined.

\section{Background}

The different phases involved in recompiling a package, program or library,
are detailed in this section. The work performed by typical compilers 
and by the DEC SRC Modula-3 compiler, possible enhancements, and related
work are discussed at key steps.

The modification time of files comprising a package are 
checked to determine which ones were modified since the last compilation.
These files always need to be recompiled.
In an integrated development environment, this information may be provided
by the editor, if all modifications are performed through it.

If any type checking, or data structure member offset computation,
is performed at compile time, files have dependencies upon imported
files containing declarations. This is not the case for
languages such as Smalltalk \cite{Smalltalk} or LISP, which defer type
checking to runtime, but applies to Java, C++ and Modula-3.
When one of the declarations used by a file is changed, that file needs
to be recompiled too. 

The use of a text level macro preprocessor, for
importing declarations from other files, makes such dependency analysis
extremely difficult in C and C++. The traditional approach embodied by
\emph{makefile}s is to recompile a file whenever an included file was 
modified. The DEC SRC Modula-3 compiler remembers the fingerprint 
of each declaration used by a module, and recompiles the module only 
when any of the declarations used has a different fingerprint. 
This much finer grain dependency analysis, 
(at the level of individual type declaration instead of complete header file), 
explains in large 
part the smaller recompilation times typically associated with more 
structured languages, as compared to C/C++.

Computing the minimal set of files needing recompilation is the smart
recompilation phase and has been studied in 
\cite{smart-first}, \cite{smart-stats}, \cite{make-stats}, 
and \cite{server-m2precomp}. Any reduction in the number of files to
recompile, due to a finer analysis, has a direct impact on the 
total recompilation time.

The set of modified files, and of files potentially affected by
modifications in files on which they depend, is recompiled in the
parsing and code generation phase.
Compiling the individual files is often the most time consuming
step. While minimizing the number of files to recompile is important,
it is also possible to reduce the number of imported interfaces to
parse, and to speedup the code generation phase.

The DEC SRC Modula-3 compiler already has a cache for imported interface
files. Each imported interface is read at most once in each recompilation.
A first extension, described in section 3, was to convert the compiler 
into a compilation server. This way, an imported file may be kept in the
interface cache and need not be read and reparsed if it did not change 
between two compilations.

Koehler and Horspool \cite{server-ccc} 
worked on a compilation server for the C language. 
Most of their effort was spent analyzing the pre-processor
context, to determine if the imported file can be reused in different
importers. 
It requires a validation scheme rather different 
than the one proposed here. 
Onodera 
proposed \cite{server-ibm} a compilation server
for a different language, COB, which has 2 kinds of interfaces,
one similar to C and one similar to Modula-3. 

The code generation time is a significant fraction of the total
compilation time. A fast code generator for Linux ELF, based on an existing
code generator for NT in SRC Modula-3, is a second extension described
in section 4.
Tanenbaum et al. \cite{back-kit}, and Fraser et al. \cite{burs-time-space}
have obtained interesting results with fast but less flexible non optimizing
backends.

In the prelink phase, a number of application wide informations 
may need to be computed, to be used by the run-time system. Non
OO languages like C had little or none such information. In Modula-3,
the initialization order of modules (based on the modules dependency graph),
the run time type information (after resolving type equivalence),
the coherence of opaque types revelations, and the structures for
checking inheritance relationships in constant time, are computed during
this prelink phase. In Java, initialization order is determined at
runtime. C++ does not specify an initialization order and only
recently started to offer run time type identification.

The final step, performed by the link editor, is assembling the 
object code of all the modules into the final executable. With
dynamic linking, such as in Linux ELF \cite{bk-abi386}, 
the amount of processing
required is greatly reduced. All references internal to a module use
position independent code and do not require further link editing.
Moreover, links to libraries are resolved at execution time. Data symbols
in libraries are resolved upon loading the executable but procedure
references are resolved only when needed, as the execution proceeds.

\section{Compilation Server}

A compilation server reduces the recompilation time by maintaining
some information across compilations, instead of reading and reparsing the
corresponding files each time. Imported interfaces represent the
bulk of the information required when compiling a file and often
do not change from one compilation to another. Thus, the purpose
of the compilation server is to maintain the interface cache across
compilations.

Implementing a compilation server for C/C++ presents 
serious difficulties because of the preprocessor mechanism.  For
instance, the \#ifdef in
Figure \ref{fig-contexte} can lead to two 
different variables in the symbol table, depending on the
value of DEBUG.  Moreover, the compilation of an included \emph{.h}
file does not lead to an independent object file.
Instead, the object code associated with it is part of 
the relocatable object files produced by the compilation
of \emph{``.c''} files that include the \emph{``.h''}
file.

\begin{figure}
  \begin{center}
    \begin{picture}(60,40)
\texttt{
\put(0,0){\shortstack[l]{
  \#ifdef DEBUG \\
  int x \\
  \#else \\
  int y  \\
  \#endif}}
}
    \end{picture}
  \end{center}
  \caption{\#ifdef example} \label{fig-contexte}
\end{figure}

In languages with explicit interfaces such as Modula-3, the content
of an interface is not dependent on the importing file and can be
reused for several importing files, even across compilations.
Some languages such as Eiffel and Java extract the interface from
the program file. When the program file is recompiled, the interface
is extracted. It could be stored in the interface cache of a compilation
server in the same way.

\begin{figure}
  \begin{center}
    \begin{picture}(180,132)

      \put(80,100){\framebox(20,20){E}}

      \put(40,100){\framebox(20,20){D}}

      \put(10,90){\dashbox{1}(120,40){}}
      \put(140,108){library 1}

      \put(100,50){\framebox(20,20){C}}

      \put(60,50){\framebox(20,20){B}}

      \put(20,50){\framebox(20,20){A}}

      \put(10,40){\dashbox{1}(120,40){}}
      \put(140,58){library 2}

      \put(60,0){\framebox(20,20){P}}

      \put(154,0){\shortstack[l]{new program \\ or library}}

      \put(70,70){\vector(2,3){20}}
      \put(70,70){\vector(-2,3){20}}
      \put(30,70){\vector(2,3){20}}

      \put(100,60){\vector(-1,0){20}}
      \put(40,60){\vector(1,0){20}}

      \put(70,20){\vector(4,3){40}}
      \put(70,20){\vector(0,1){30}}
      \put(70,20){\vector(-4,3){40}}

    \end{picture}
  \end{center}
  \caption{Dependencies among imported Modula-3 interfaces} \label{fig-imports}
\end{figure}
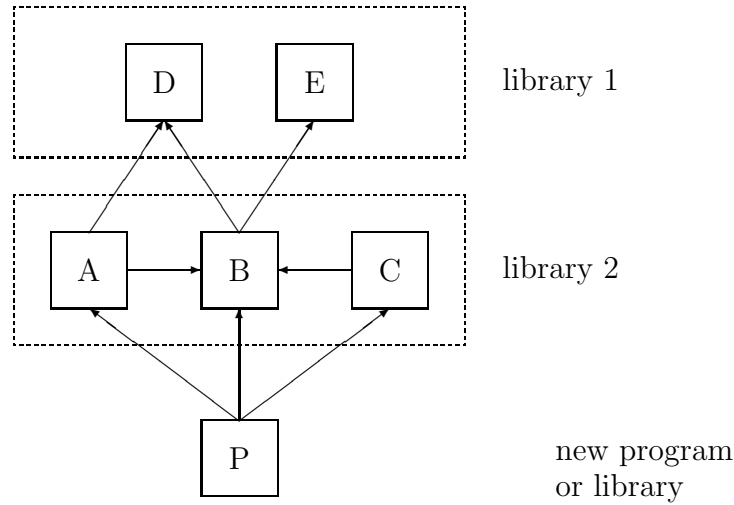

Figure \ref{fig-imports} shows the dependencies
between interfaces A to E used by a program P.  An
arrow shows that a module or an interface imports
another interface.  This
acyclic directed graph imposes a compilation
order.  The interfaces at the leaves are compiled first.
Each interface is compiled seperately, i.e. it has
its own associated relocatable object file. However, 
an interface would normally need to be parsed again whenever
it is imported.  This is the costly part of the 
compilation avoided with the interface cache.
The parsing result, the abstract syntax
tree (\emph{AST}), is stored in the interface cache.

Between compilations, some of the interfaces in the cache
may become invalid. Indeed, if the associated
source file has been modified or if a directly or
indirectly imported interface is invalid, the interface is declared
invalid. In
the last case, the fine grain dependency analysis is used to
determine if any declaration actually used by the interface has changed.
If not, the interface is still valid. Otherwise, the interface is removed
from the cache and will need to be read and parsed again.

Thus, the
validation algorithm is to recursively visit each
imported interface in the graph to see if its associated source
file has been modified or if imported declarations have changed.  
A time stamp is used to mark each valid interface
as valid for this recompilation. When the same interface appears in the
import graph of another file, the time stamp indicates that it has already
been checked as valid.
The existing implementation of DEC SRC Modula-3 did not have a validation
phase since the interface cache was not kept across compilations.
Any interface in the cache was necessarily parsed in the current 
compilation and therefore still valid.

A further optimization is added for interfaces imported from 
separate libraries (\emph{packages}).
Indeed, each \emph{package} has an associated file containing the
information used by the smart recompilation system. Any modification
to an interface in a library, and subsequent recompilation, changes the
modification time of the associated file. Thus, the modification time
of library interfaces is checked only if the associated file has changed.
Many libraries being seldom changed, this simple optimization avoids a
large number of file modification time checks. 
The generic interfaces (the equivalent of the
templates in C++) are not put in the cache
because they don't have a corresponding AST. Their instanciations, however,
are eligible to be cached.

Figures \ref{fig-compil-1} and \ref{fig-compil-2}
show the general structure of the compiler before and
after being transformed into a server. Libraries are represented as
ovals and programs as rectangles.
Import relationships are indicated by arrows.
Programs executing other programs are shown with dashed lines.

\begin{figure}
  \begin{center}
    \begin{picture}(260,100)

      \put(0,80){\framebox(36,12){m3build}}

      \put(68,86){\oval(36,12)}
      \put(52,84){m3quake}

      \put(150,80){\framebox(36,12){m3}}

      \put(130,56){\oval(40,12)}
      \put(118,54){m3linker}

      \put(160,36){\oval(40,12)}
      \put(148,34){m3front}

      \put(210,50){\oval(40,12)}
      \put(198,48){m3back}

      \put(200,14){\oval(46,12)}
      \put(186,12){m3middle}

      \put(100,0){\dashbox(140,100)[b]}
      \put(104,4){``driver''}

      \put(36,86){\vector(1,0){13}}
      \multiput(87,86)(8,0){7}{\line(2,0){4}}
      \put(144,86){\vector(1,0){6}}

      \put(168,80){\vector(-2,-1){36}}
      \put(168,80){\vector(-1,-4){9.6}}
      \put(168,80){\vector(1,-2){29.8}}
      \put(168,80){\vector(3,-2){36}}

      \put(160,30){\vector(4,-1){38.4}}
      \put(210,44){\vector(-1,-2){12}}

    \end{picture}
  \end{center}
  \caption{Structure of the SRC Modula-3 compiler}\label{fig-compil-1}
\end{figure}
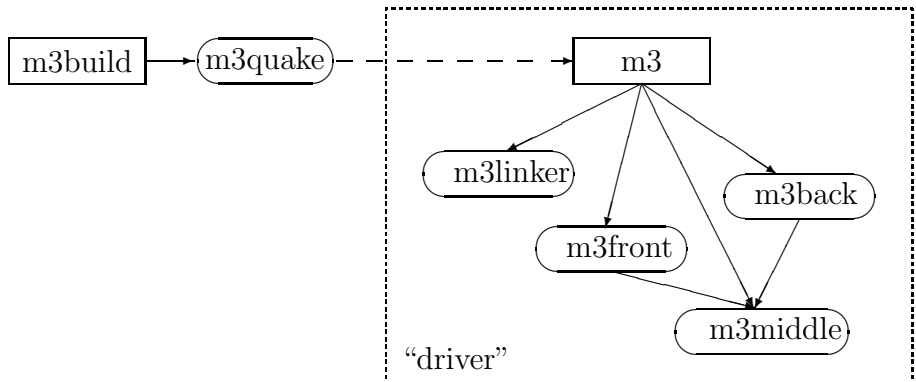

\begin{figure}[t]
  \begin{center}
    \begin{picture}(210,116)

      \put(0,102){\framebox(36,12){m3client}}

      \put(0,72){\framebox(36,12){m3server}}

      \put(78,78){\oval(38,12)}
      \put(62,76){m3quake}

      \put(144,78){\oval(40,12)}
      \put(142,76){m3}

      \put(110,48){\oval(40,12)}
      \put(95,46){m3linker}

      \put(140,28){\oval(40,12)}
      \put(128,26){m3front}

      \put(190,42){\oval(40,12)}
      \put(178,40){m3back}

      \put(180,6){\oval(46,12)}
      \put(164,4){m3middle}

      \multiput(18,99)(0,-2){4}{\circle*{1.4}}

      \put(36,78){\vector(1,0){23}}
      \put(97,78){\vector(1,0){27}}

      \put(148,72){\vector(-2,-1){36}}
      \put(148,72){\vector(-1,-4){9.6}}
      \put(148,72){\vector(1,-2){29.8}}
      \put(148,72){\vector(3,-2){36}}

      \put(140,22){\vector(4,-1){38.4}}
      \put(190,36){\vector(-1,-2){12}}

    \end{picture}
  \end{center}
  \caption{Structure of the modified SRC Modula-3 compiler}\label{fig-compil-2}
\end{figure}

Packages \emph{m3front} and \emph{m3back} are
the frontend and the backend respectively. \emph{M3linker}
is the smart recompilation system.  Package
\emph{m3} contains the main procedure of the compiler.
\emph{M3quake} is a simple interpreter that parses
the m3makefiles and passes parameters to the 
compiler (\emph{m3}). \emph{M3build} initiates the \emph{m3quake}
interpreter in the appropriate directory, inserts the platform and
package dependent definitions, and sends the m3makefile to \emph{m3quake}. 
The connections to the two available code generators, and to the linker,
were left out for simplicity.

The existing compiler involved several processes which were merged in a
single executable program calling upon several libraries.
Packages \emph{m3} and \emph{m3quake} were converted to libraries.
\emph{M3build} acted as the remaining program and, instead of exiting
after a recompilation, awaited further commands from clients using
network objects \cite{doc-netobj}, as shown in Figure \ref{fig-netobj}.

Each client request consists in a package name and location,
build options if any, and a network output stream to receive the
error messages. 
The same compilation server can process requests for different
packages. Simultaneous recompilation requests are serialized.

\begin{figure}
  \begin{center}
    \begin{picture}(296,94)
\texttt{
\put(0,0){\shortstack[l]{INTERFACE M3Server; \\ \\
IMPORT NetObj, Thread, Wr, Pathname, TextList; \\ \\
EXCEPTION \\
\ \ Error (TEXT); \\ \\
TYPE \\
\ \ T = NetObj.T OBJECT METHODS \\
\ \ \ \ compile(init\_dir:Pathname.T;options:TextList.T;writer:Wr.T) \\
\ \ RAISES {NetObj.Error, Thread.Alerted, Error}; \\
\ \ END; \\ \\
END M3Server.}}}
    \end{picture}
  \end{center}
  \caption{Network object interface to the compilation server}\label{fig-netobj}
\end{figure}

A small program, \emph{m3client}, acts as a client that passes 
compilation requests to the server. It replaces the compilation command
normally entered from the command line or through the editor menus.

There is currently no mechanism or strategy to remove interfaces
from the cache (e.g. least recently used). This is not a problem for
a few users working on a few packages. However, if the compilation server
executes for weeks and new packages are constantly being added, the
memory growth is likely to become a problem. Deciding on an efficient
strategy may require some study but would be a minor implementation
effort. The problem of multi-user access control has not been addressed
either. Sharing a compilation server would bring interesting benefits only
in very specific environments.

An alternative to maintaining the parsed interfaces in memory in a server
process across compilations, is to store on disk pre-parsed versions 
of the interfaces. This would be faster than having no compilation
server, if writing these pre-parsed interfaces is much quicker than
the time saved by not having to re-parse the interfaces. However, this
is slower than keeping the parsed interfaces already in memory, provided
that enough memory is available.

\section{Fast Code Generation}

Portable, retargetable, optimizing compilers such as gcc
are built in several layers and construct abstract syntax trees,
an intermediate language representation, and an assembly language
file at different stages during the
compilation. Furthermore, the interface between the Modula-3 frontend,
written in Modula-3, and the gcc based backend introduces another 
intermediate representation.

While the gcc based backend is retained to benefit from its
optimizing capabilities and wide range of supported platforms,
an integrated code generator may be used on some
platforms for faster compilation during the edit compile debug 
cycle. The code generator is fed by the frontend with simple
virtual machine instructions and generates object code directly.
It cannot perform sophisticated optimizations or target multiple
platforms.

An existing code generator for NT under Intel 386 was used as a base.
It was modified to support Linux ELF object files \cite{bk-abi386}, 
and to produce position independent code and debugging information. 
Position independent
code allows efficient dynamic linking, another important ingredient
for fast recompilation. Debugging information generation increases slightly
the code generation time but is almost essential for adequately
supporting the edit compile debug cycle.

Figure \ref{fig-LangInter} \emph{a)} shows a
simple Modula-3 procedure returning the sum of
two values received as arguments.  In \emph{b)},
the sequence of methods calls issued by the Modula-3 frontend,
and implemented by the code generator object, is illustrated.
The methods supported by the code generation object are sometimes called
the intermediate language. The arguments for the methods
were omitted in the example.  In \emph{c)},
the machine code produced by the backend is presented
in the AT\&T assembly language format for Intel 386 Architecture.

\begin{figure}[t]
  \begin{center}
    \begin{picture}(200,200)
       \texttt{
\put(0,166){\shortstack[l]{PROCEDURE Add(i, j: INTEGER): INTEGER \(=\)\\
\ \ BEGIN  \\
\ \ \ \ RETURN i \(+\) j;  \\
\ \ END Add;}}}

\put(80,154){(a)}

\texttt{
\put(0,137){begin\_procedure(\ldots)}
\put(0,81){\shortstack[l]{load(\ldots)  \\
load(\ldots)  \\
add(\ldots)}}
\put(0,54){\shortstack[l]{exit\_proc(\ldots) \\
end\_procedure(\ldots)}}
}

\put(32,0){(b)}

\put(84,140){\line(1,0){20}}
\put(104,143){\line(0,-1){49}}

\put(40,105){\line(0,-1){28}}
\put(40,85){\line(1,0){64}}
\put(104,75){\line(0,1){16}}

\put(58,67){\line(1,0){46}}
\put(104,70){\line(0,-1){6}}

\put(74,56){\line(1,0){30}}
\put(104,61){\line(0,-1){45}}
\texttt{
\put(106,96){\shortstack[l]{pushl  \%ebp  \\
movl   \%esp,\%ebp  \\
pushl  \%ebx  \\
pushl  \%esi  \\
pushl  \%edi}}
\put(106,18.4){\shortstack[l]{
movl   0x8(\%ebp),\%edx  \\
addl   0xc(\%ebp),\%edx  \\
movl   \%edx,\%eax  \\
popl   \%edi  \\
popl   \%esi  \\
popl   \%ebx  \\
leave \\
ret   }}}

\put(126,0){(c)}

    \end{picture}
  \end{center}
  \caption{Compilation of a simple procedure.} \label{fig-LangInter}
\end{figure}

\begin{figure}
  \begin{center}
    \begin{picture}(200,84)

      \put(50,0){\framebox(90,80)}
      \put(50,70){\line(1,0){90}}
\thicklines
      \put(50,60){\line(1,0){90}}
\thinlines
      \put(50,50){\line(1,0){90}}
      \put(50,40){\line(1,0){90}}
      \put(50,30){\line(1,0){90}}
      \put(50,20){\line(1,0){90}}
      \put(50,10){\line(1,0){90}}

      \put(72,72.6){argument 1 (\texttt{j})}
      \put(72,62.6){argument 0 (\texttt{i})}
      \put(72,52.6){ return address}
      \put(76,42.6){previous \texttt{ebp}}
      \put(68,32.6){temporary variable}
      \put(90,22.6){\texttt{ebx}}
      \put(90,12.6){\texttt{esi}}
      \put(90,2.6){\texttt{edi}}

      \put(15,72.6){C(\%\texttt{ebp})}
      \put(15,62.6){8(\%\texttt{ebp})}
      \put(15,52.6){4(\%\texttt{ebp})}
      \put(15,42.6){0(\%\texttt{ebp})}
      \put(13,32.6){-4(\%\texttt{ebp})}
      \put(13,22.6){-8(\%\texttt{ebp})}
      \put(13,12.6){-C(\%\texttt{ebp})}
      \put(10,2.6){-10(\%\texttt{ebp})}

      \put(144,76){high addresses}
      \put(144,0){low addresses}

    \end{picture}
  \end{center}
  \caption{Stack frame} \label{fig-stack}
\end{figure}
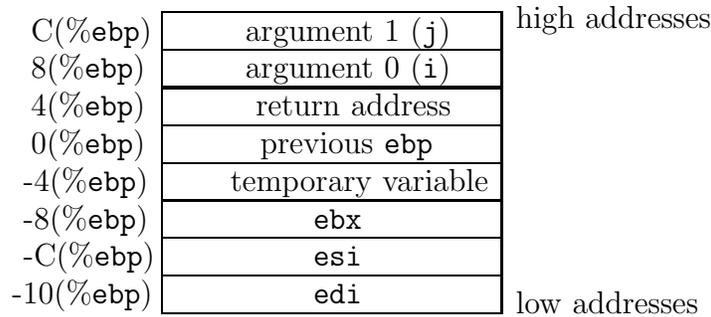

At the beginning of a procedure, 
the method \emph{begin\_procedure}
is called by the frontend, and the 6 first machine
instructions are generated. This saves on the stack and initializes the
registers used
for setting up a new frame according to the calling convention.
Figure \ref{fig-stack} shows the structure of the stack frame
constructed with these instructions.  The \emph{ebp}
register (base register) is used to reference
other values in the stack frame by specifying a
4 byte offset in an indexed address mode.
The stack frame also includes
the arguments to the procedure, the return
address of the calling procedure, 
the value of \emph{ebp}
for the previous stack frame,
local variables,
and some saved register values.

Thereafter, the machine instructions corresponding to 
the body of the Modula-3 procedure are generated.
The \emph{load} method is called twice by the frontend, once
for each operand. No machine code is generated for this.
Instead, the operands are
pushed on the operand stack in the data structures maintained by
the code generator. The call of the method \emph{add}
by the frontend generates the code for adding the operands previously
pushed on the stack. On the i386, at least one operand
of \emph{add} must be in a register. Thus, one variable must be moved from
its position in the stack frame in memory to a register
(here \emph{edx}).  The other operand can be addressed
directly in memory by instruction \emph{add}.  
The result of the addition must be returned to the calling
procedure in register \emph{eax}, as specified in the calling convention.
Thus, the value in \emph{edx} is moved to \emph{eax} 
when method \emph{exit\_procedure}
is called by the frontend.  
The call to method
\emph{end\_procedure} generates the code required
to remove the stack frame and restore affected registers.

Several optimizations to the generated code would be possible.
For example, the use of register
\emph{edx} could be avoided by using \emph{eax}
instead.  This would obviate the need to transfer the return value
to \emph{eax} in the end.
Registers \emph{ebx}, \emph{esi}
and \emph{edi} were needlessly saved and restored on the stack 
even if they were not used during the body of the 
procedure. Even though the fast code generator does not perform such
optimizations, the generated code is still more efficient than
the naive code produced by gcc without optimization, as seen in
section 5.

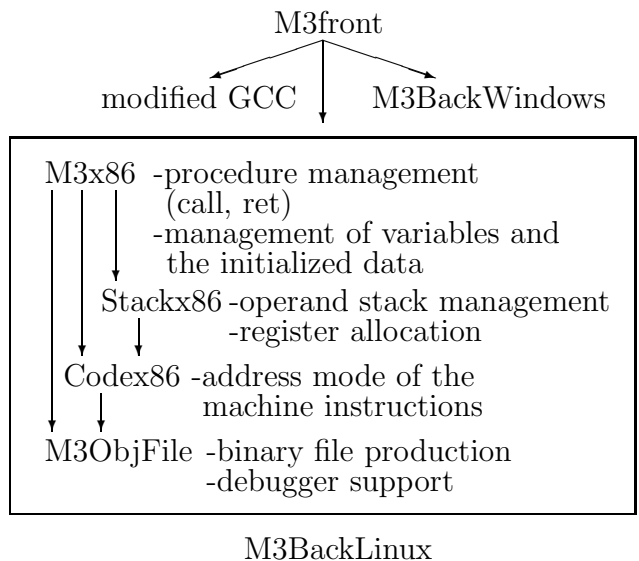
\begin{figure}
  \begin{center}
    \begin{picture}(168,148)

      \put(70,140){M3front}
      \put(83,138){\vector(0,-1){22}}
      \put(83,138){\vector(3,-1){30}}
      \put(83,138){\vector(-3,-1){30}}

      \put(24,120){modified GCC}
      \put(96,120){M3BackWindows}

      \put(0,12){\framebox(166,100)}
      \put(9,100){M3x86}
      \put(38,100){-procedure management}
      \put(41,92){(call, ret)}
      \put(38,84){-management of variables and}
      \put(41.4,76){the initialized data}

      \put(24,66){Stackx86}
      \put(58,66){-operand stack management}
      \put(58,58){-register allocation}

      \put(14,46){Codex86}
      \put(48,46){-address mode of the}
      \put(51,38){machine instructions}

      \put(9,26){M3ObjFile}
      \put(52,26){-binary file production}
      \put(52,18){-debugger support}

      \put(62,0){M3BackLinux}

      \put(11,98){\vector(0,-1){64}}
      \put(19,98){\vector(0,-1){44}}
      \put(28,98){\vector(0,-1){24}}
      \put(34,64){\vector(0,-1){10}}
      \put(24,44){\vector(0,-1){10}}

    \end{picture}
  \end{center}
  \caption{Structure of the code generator.} \label{fig-CGStruct}
\end{figure}

The code generator structure is designed around
four major objects, as shown in Figure 
\ref{fig-CGStruct}.  The main object is
\emph{M3x86} which implements
code generation for procedure
calls and returns, and produces
the global variables section. It also coordinates
the code generation performed by the other objects.
\emph{Stackx86} implements the operand stack, performs the register
allocation and generates the code to move operands between the stack
and the registers. Detailed machine instructions layouts, including
addressing modes, are obtained through the \emph{Codex86}
object. Everything is formatted into an ELF binary file, with debugging
information, by \emph{M3ObjFile}.
The relationships between these objects are represented by arrows 
in Figure \ref{fig-CGStruct}.

%
%

\section{Results}

The test set used for the performance evaluation is described in
Table \ref{tab-list}. It contains programs and libraries of different sizes,
and importing different types of libraries. Large graphical applications 
such as webscape, postcard, and ps2html tend to use large libraries, 
and involve numerous imported interfaces. Webvbt is one of the libraries
used by webscape, and columns is a smaller graphical
application. Netobjd is a small program involving network objects, and
m3browser is a module/type browser with a web server interface.
M3tohtml is a small program converting Modula-3 modules to html, and
m3front is a library implementing the Modula-3 compiler frontend.

For each package are shown the source code size, number of lines, 
number of non blank lines, number
of directly and indirectly imported interfaces, and
the memory consumed by the interface cache to represent the AST.
All tests were performed on a 75MHz Intel Pentium with 32MB of main memory
and running Linux 2.0. All times are in elapsed seconds on a single user
machine, and thus account for input/output delays.

\begin{table}
  \begin{center} 
\  \\ \ \\
    \begin{tabular}{|l||c|c|c|c|c|c|c|} \hline
package  & & & & & & imported & memory \\
compiled & size & lines & lines* & interf. & 
modules  & interfaces & required  \\ \hline \hline

columns    &  47.17K &  1553 &  1306 &   6 &   7 &  30 & 2.26M \\ \hline
netobjd    &   3.87K &   151 &   128 &   1 &   2 &  27 & 1.12M \\ \hline
webscape   &  11.72K &   374 &   347 &   0 &   1 &  69 & 2.96M \\ \hline
m3browser  & 210.58K &  7005 &  6312 &  12 &  13 &  67 & 1.93M \\ \hline
webvbt     & 194.76K &  6254 &  5364 &  21 &  20 & 120 & 3.63M \\ \hline
m3tohtml   &  83.08K &  3035 &  2656 &   8 &   9 &  35 & 1.26M \\ \hline
m3front    &  1.367M & 45827 & 39789 & 175 & 171 &  38 & 3.06M \\ \hline
postcard   & 341.81K & 10418 &  9575 &  12 &  11 & 161 & 3.93M \\ \hline
ps2html    & 315.85K & 13468 &  9197 &  30 &  30 & 111 & 3.57M \\ \hline
    \end{tabular}
\  \\  \end{center}
  \caption{Packages used to evaluate the performance}  
  \label{tab-list}
\end{table}

The performance of both the integrated
backend and the compilation server were evaluated and compared to the
original DEC SRC Modula-3 compiler version 3.6.
The compilation times for the original compiler are in Table
\ref{std-m3cgc1-reg}.

Detailed measures were obtained for the various compilation phases using the
compiler built-in timers.
Column M3$\Rightarrow$I.L is the time required to translate Modula-3
code into intermediate language, which is performed by the frontend.
I.L.$\Rightarrow$ass. is the time needed by the gcc based backend (m3cgc1)
to convert
the intermediate language into assembly, and
ass.$\Rightarrow$reloc. is the time taken by
the assembler to produce a relocatable binary file.

\begin{table}
  \begin{center}\  \\ \  \\
    \begin{tabular}{|l||c|c|c|c|c|c|c|} \hline
\multicolumn{1}{|c||}{ } & \multicolumn{7}{c|}{Time (in seconds)} \\ \cline{2-8}
package & smart & M3 $\Rightarrow$ & I.L. $\Rightarrow$ & ass. 
$\Rightarrow$ &  & & \\ 
compiled & recomp. & I.L. & ass. & reloc. & linking & other & total \\
\hline \hline

columns    & 1.19 &  4.21 &   8.65 &   4.44 &  0.85 &   0.32 &  19.66  \\ \hline
netobjd    & 0.78 &  0.87 &   1.50 &   0.85 &  0.43 &   0.32 &   4.75  \\ \hline
webscape   & 3.48 &  2.56 &   4.06 &   1.51 &  1.50 &   0.91 &  14.02  \\ \hline
m3browser  & 1.04 & 10.08 &  36.09 &  11.88 &  1.11 &   0.81 &  61.01  \\ \hline
webvbt     & 4.80 & 14.17 &  37.52 &  16.83 &  0.97 &   2.03 &  76.32  \\ \hline
m3tohtml   & 0.79 &  4.00 &  16.49 &   6.39 &  0.85 &   0.61 &  29.13  \\ \hline
m3front    & 1.62 & 69,56 & 220.36 & 108.03 &  5.84 &  17.32 & 422.73  \\ \hline
postcard   & 2.79 & 22.38 &  55.34 &  16.43 &  4.16 &   0.74 & 101.84  \\ \hline
ps2html    & 3.43 & 28.24 &  84.38 &  27.34 &  6.02 &   1.48 & 150.89  \\ \hline
    \end{tabular} \  \\
  \end{center}
  \caption{Compilation with the gcc based code generator}  
  \label{std-m3cgc1-reg}
\end{table}

Table \ref{back-opt-reg} presents the results for
the compiler with the integrated backend instead of
the gcc based backend.  
The fast code generator
goes directly from the code generating method calls to the relocatable 
binary file,
as indicated by
M3$\Rightarrow$reloc. Since the code generation is performed at the same
time as parsing, the time required for each phase is not available
separately. 

The compilation time reduction is significant for all packages.  
When the compilation
time is larger than 20 seconds with the gcc based backend, the 
total compilation time may be cut in half with
the integrated backend. Other tests \cite{Collin1997}
demonstrate that the production of position 
independent code does not significantly affect the compilation time.
However, the generation of debugging information
increases the total compilation time by 10-30\%.

To evaluate the quality of the generated code, a program
was compiled with the gcc based backend without optimization
and with full optimization (-O2), and with the fast code generator.
The execution time for the version compiled with the fast
code generator was 6\% faster than without optimization and
9\% slower than with optimization.
The memory footprint of the program compiled
with the fast code generator is 18\% smaller than without optimization
and 14\% larger than with optimization.

\begin{table}
  \begin{center} \  \\ \  \\
    \begin{tabular}{|l||c|c|c|c|c|c|c|} \hline
\multicolumn{1}{|c||}{ } & \multicolumn{5}{c|}{Time (in seconds)} \\ \cline{2-6}
package & smart & M3 $\Rightarrow$ &   & & \\ 
compiled & recomp. & reloc. & linking & other & total \\
\hline \hline

columns    & 1.16 &  6.46 &  0.85 &  0.61 &   9.08  \\ \hline
netobjd    & 0.77 &  1.24 &  0.43 &  0.39 &   2.83  \\ \hline
webscape   & 3.98 &  2.93 &  1.69 &  1.14 &   9.74  \\ \hline
m3browser  & 0.97 & 14.91 &  0.85 &  0.46 &  17.19  \\ \hline
webvbt     & 4.52 & 21.60 &  1.01 &  2.02 &  29.15  \\ \hline
m3tohtml   & 0.94 &  5.83 &  1.06 &  0.39 &   8.22  \\ \hline
m3front    & 2.45 & 95.54 &  8.05 & 16.88 & 122.92  \\ \hline
postcard   & 3.07 & 31.75 &  8.51 &  1.09 &  44.42  \\ \hline
ps2html    & 3.62 & 35.73 &  4.77 &  1.28 &  45.40  \\ \hline

    \end{tabular} \  \\ 
  \end{center}
  \caption{Compilation with the integrated backend}  
  \label{back-opt-reg}
\end{table}

These results are consistent with those reported by
Tanenbaum and al. \cite{back-kit}, where they obtained 
a speedup between 2 and 3 by using a simplified backend.
Their backend however still retained the ability to
interface to different frontends and targets.

The compilation server was evaluated both under a full recompilation
and in a typical situation where only a few files were modified.
For the first case, the executable and all object files were removed
before recompiling. This is shown in Table
\ref{serv-netobj}.

The savings brought by maintaining the interface cache
across two compilations are smaller than originally anticipated.
Only small gains are obtained for the small packages.
Since the SRC Modula-3 compiler already contains an interface cache,
the time to parse the interfaces is relatively small as compared to the
code generation time, and a small fraction of the time is saved by 
caching the interfaces across compilations.

More disturbing is the slight degradation obtained for some of the 
larger packages.
Removing the network objects communication between the client and
server did not
change the results. The explanation lies in the increased memory footprint
of the compilation server. With 32 MB, there is some competition for
physical memory between
buffered files (e.g. libraries, object files and the generated executable), 
and executing programs (e.g. the compiler and linker). Indeed, when the
same tests were repeated on a computer with twice as much physical memory,
a slight improvement was measured instead of a degradation.

\begin{table}
  \begin{center} \  \\ \  \\
    \begin{tabular}{|l||c|c|c|c|c|c|c|} \hline
\multicolumn{1}{|c||}{ } & \multicolumn{7}{c|}{Time (in seconds)} \\ \cline{2-8}
package & smart & M3 $\Rightarrow$ &I.L. $\Rightarrow$ & ass. 
$\Rightarrow$ & & & \\ 
compiled & recomp. & I.L. & ass. & reloc. & linking & other & total \\
\hline \hline

columns    & 0.63 &  1.96 &   9.04 &   4.45 &  0.86 &  0.26 &  17.20  \\ \hline
netobjd    & 0.46 &  0.28 &   1.52 &   1.09 &  0.44 &  0.25 &   4.04  \\ \hline
webscape   & 3.00 &  1.58 &   4.06 &   1.55 &  1.71 &  0.65 &  12.55  \\ \hline
m3browser  & 0.63 &  8.90 &  37.19 &  12.24 &  1.55 &  0.40 &  60.91  \\ \hline
webvbt     & 4.46 & 12.04 &  39.80 &  17.41 &  1.29 &  2.31 &  77.31  \\ \hline
m3tohtml   & 0.41 &  3.29 &  16.40 &   6.90 &  0.86 &  0.37 &  28.23  \\ \hline
m3front    & 1.39 & 77.84 & 235.90 & 121.02 & 34.70 & 31.18 & 502.03  \\ \hline
postcard   & 2.48 & 21.67 &  57.73 &  18.35 &  7.32 &  0.72 & 108.27  \\ \hline
ps2html    & 3.12 & 24.73 &  89.48 &  30.44 & 11.54 &  1.59 & 160.90  \\ \hline

    \end{tabular}\  \\
  \end{center}
  \caption{Full recompilation of the packages with AST's in the cache}  
  \label{serv-netobj}
\end{table}

In a full recompilation, the existing interface cache brings most of the
benefits, and the code generation phase largely dominates because all modules
need to be recompiled. Therefore, the savings brought by the server are
minor as compared to the total compilation time and may even turn into
a degradation if memory is short. The picture improves however when
the compilation server is combined with the fast code generator.
Indeed, the 2.46s reduction for columns represents 12.5\% of
19.66s but amount to 27.1\% of 9.08s.
These savings
only involve the smart recompilation and frontend parsing phase, and
are independent of the code generation time.

The tests presented in Tables \ref{partiel-1} and
\ref{partiel-2} are more typical of the edit compile debug cycle.
They consist in recompiling a large graphical application, postcard,
after 2 files, and 4 files, are modified. Four different cases are
covered: existing compiler and code generator, fast code generator,
compilation server, compilation server with fast code generator.
For these tests, the computer used had the same characteristics except
a Pentium Pro 180MHz processor and 64MB of RAM.

\begin{table}
  \begin{center} \  \\ \  \\
    \begin{tabular}{|l||c|c|c|c|c|c|c|} \hline
\multicolumn{1}{|c||}{ } & \multicolumn{7}{c|}{Time (in seconds)} \\ \cline{2-8}
compiler & smart & M3 $\Rightarrow$ & I.L. $\Rightarrow$ & ass. 
$\Rightarrow$ &  & & \\ 
used & recomp. & I.L. & ass. & reloc. & linking & other & total \\
\hline \hline

standard       &      &      &      &      &      &      &       \\
with m3cgc1    & 0.73 & 0.97 & 2.73 & 0.86 & 0.42 & 0.22 & 5.93  \\ \hline
server         &      &      &      &      &      &      &       \\
with m3cgc1    & 0.76 & 0.62 & 2.73 & 0.84 & 0.42 & 0.15 & 5.52  \\ \hline
standard       &      & \multicolumn{3}{c|}{    } &      &      &       \\
with integrated   & 0.74 & \multicolumn{3}{c|}{1.84} & 0.42 & 0.24 & 3.24  \\ \hline
server        &      & \multicolumn{3}{c|}{    } &      &      &       \\
with integrated   & 0.64 & \multicolumn{3}{c|}{1.36} & 0.42 & 0.20 & 2.62  \\ \hline

    \end{tabular}\  \\
  \end{center}
  \caption{Recompilation of postcard after modifications to 2 files}  
  \label{partiel-1}
\end{table}

\begin{table}
  \begin{center}\  \\ \  \\
    \begin{tabular}{|l||c|c|c|c|c|c|c|} \hline
\multicolumn{1}{|c||}{ } & \multicolumn{7}{c|}{Time (in seconds)} \\ \cline{2-8}
compiler & smart & M3 $\Rightarrow$ & I.L. $\Rightarrow$ & ass. 
$\Rightarrow$ &  & & \\ 
used & recomp. & I.L. & ass. & reloc. & linking & other & total \\
\hline \hline

standard       &      &      &      &      &      &      &       \\
with m3cgc1    & 0.75 & 1.89 & 4.83 & 1.68 & 0.42 & 0.20 & 9.77  \\ \hline
server         &      &      &      &      &      &      &       \\
with m3cgc1    & 0.44 & 1.09 & 4.85 & 1.68 & 0.43 & 0.17 & 8.66  \\ \hline
standard       &      & \multicolumn{3}{c|}{    } &      &      &       \\
with integrated   & 0.73 & \multicolumn{3}{c|}{2.66} & 0.43 & 0.27 & 4.09  \\ \hline
server         &      & \multicolumn{3}{c|}{    } &      &      &       \\
with integrated   & 0.51 & \multicolumn{3}{c|}{2.05} & 0.42 & 0.24 & 3.22  \\ \hline

    \end{tabular}\  \\
  \end{center}
  \caption{Recompilation of postcard after modifications to 4 files}  
  \label{partiel-2}
\end{table}

These tests clearly demonstrate that the compilation server and fast
code generator contribute independently to the recompilation time
reduction. Their combined effect in that case amounts to a reduction
from 9.77s to 3.22s. This is especially impressive considering that
the Modula-3 compiler is already much more efficient than most C/C++
compilers, because of the interface cache and fine grain dependency
analysis allowed by the structured interface mechanism.

%
%

\section{Conclusion}

A fast recompilation system for a modular, compiled, object oriented
language was presented. It benefits from the existing interface cache
and fine grain dependency analysis, and was extended with a persistent
interface cache, fast code generator, and Linux ELF dynamic libraries
support. The resulting recompilation time reduction is impressive.
For a large graphical application, recompiling after a few files were
modified took 3.22s instead of 9.77s.

In the coming years, faster processors, larger main memory and
more complex programs may be expected. The average size
of each file, and the number of files modified between compilations
are not expected to change significantly however. The net effect
would be a gradual reduction of the file parsing and code generation 
time, which currently dominates the recompilation time at 2.05s out of
3.22s. The smart recompilation and linking time accounts for much of
the remaining time (.92s out of 1.17s). These may be adversely affected
by the increasing programs complexity and eventually dominate the
recompilation time.

Increases in programs complexity are likely to change more the number of 
libraries imported by each program rather than the size of
each program file or library. The smart
recompilation system could accordingly be further optimized in several ways.
More information about each imported library could be cached, as applications
importing more libraries have much larger times for the smart recompilation
phase. A tighter integration between the program editor and the compilation
server could allow the smart recompilation to perform most of its
work incrementally. When a file is opened and modified, the compiler could
pre-compute which files are affected, and need to be recompiled or
dependency checked. When a file is saved, the recompilation of that file
and affected files may proceed. Thus, when the last file is saved,
only that last file, and the files affected, would need recompilation.

The final recompilation steps, prelinking and linking, 
involve the complete program and may therefore
become the dominant step as program complexity increases.
Interestingly, the Linux ELF linker is surprisingly efficient for
dynamically linked libraries. Only a small fraction of each library needs
to be read by the linker. Yet, because of the lazy procedure linking
algorithm used in ELF, the program startup penalty imposed by dynamic
linking is small. As may be seen from the results, the linking time
is mostly affected by the package size rather than by 
the number of imported libraries.

Linking is an I/O intensive process, and the availability of free RAM for
the I/O buffer cache strongly impacts the elapsed time. This was apparent
for large programs where the compilation server actually 
degraded the performance
because of the competition for memory. Assuming sufficient physical memory,
it may be possible to reduce the linking time by dynamically loading each
object file as it is being recompiled. This would remove the linking step
which reads all object files from disk, merges the files into an executable,
and writes the executable to disk, before loading the executable.

The prelinking step, listed as \emph{other}, currently uses a small
fraction of the overall recompilation time. It checks the
coherency of opaque types revelations, determines the modules
initialization order, and generates the run time type identification
data structures. The processing time is proportional to the total
number of modules in the final program. With increasing program
complexities, this may in the long term become an important factor.

In C++, run time type identification is a recent addition and is
likely to bring different problems. Indeed, a declaration appearing in
a \emph{.h} file may be included and compiled several times. The prelinker
must ideally remove duplicate virtual methods tables and run time type
information. The Java language does not specify a static initialization 
order nor does it have opaque types to check. The prelinker step is thus 
mostly avoided.
However, since a class must be initialized at its first
active use, many tests must be inserted at run time to initialize classes
when they are used, if it is the first time. Thus, the savings in the prelink
phase are offset by the run time overhead.

The fast code generator is now part of the freely redistributable
Polytechnique Montreal Modula-3 distribution, originally found at
http://m3.polymtl.ca/\-m3/\-pkg but now hosted at 
http://www.elegosoft.com/\-pm3/,
and received enthusiastic feedback from users around the world.
The compilation server is available separately at
http://www.professeurs.polymtl.ca/\-michel.\-dagenais/\-pkg/\-m3server\-.tar\-.gz.

\section*{Acknowledgments}

The financial support of the National Sciences and Engineering Research
Council of Canada, as a scholarship to the first author, while he was a
master's student at Ecole Polytechnique de Montreal, and an operating grant
to the second author, is gratefully acknowledged.
This work would not have been possible without the excellent environment
provided by the DEC SRC Modula-3 compiler. Bill Kalsow, Allan Heydon,
and Greg Nelson contributed to this project through numerous technical 
discussions.

\bibliographystyle{ieeetr}
\bibliography{papers,books,docs_net}

\end{document}